\documentstyle[preprint,epsf]{jpsj}
\recdate{January 26, 1998}
\title{
Raman Scattering in the Inorganic Spin-Peierls System $\alpha '$-Na$_{1-\delta}$V$_2$O$_5$
}
\author{
Haruhiko {\sc Kuroe}, Hideki {\sc Seto}, Jun-ichi {\sc Sasaki}, Tomoyuki {\sc Sekine}, \\
Masahiko {\sc Isobe}$^{1}$ and Yutaka {\sc Ueda}$^{1}$
}
\inst{
Department of Physics, Sophia University, 7-1 Kioi-cho, Chiyoda-ku, Tokyo 102\\
$^{1}$Institute for Solid State Physics, The University of Tokyo, 7-22-1 Roppongi, Minato-ku, Tokyo 106
}
\abst{
	We have studied the spin-Peierls (SP) transition in $\alpha '$-Na$_{1-\delta}$V$_2$O$_5$ ($\delta$ = 0, 0.01 and 0.1) by means of Raman scattering.
	At room temperature, we observe six $A_1$ and three $A_2$ phonon modes and a broad Raman band.
	Below $T_{\rm SP}$ several new peaks and a new broad band appear.
	The new peak at 62 cm$^{-1}$ originates from the SP-gap excitation.
	The new peak at 128 cm$^{-1}$ and the new broad band between 130 and 400 cm$^{-1}$ come from two magnetic excitations.
	The new peaks at 102, 646 and 944 cm$^{-1}$ are assigned to the folded phonon modes and their Na$^{+}$-ion deficiency dependence shows that the defect of Na$^{+}$ ion suppresses the SP transition.
	The polarized Raman spectra below $T_{\rm SP}$ suggest that the possible crystal symmetry is $C_{s}^{2}(Pn)$ or $C_{1}^{1}(P1)$.
	The asymmetric lineshape of the 531-cm$^{-1}$ peak superimposed on the electronic Raman band from the $d$-$d$ transition around 600 cm$^{-1}$ is interpreted in terms of the Fano resonance between the electronic continuous band and the phonon with a finite lifetime.
	The defects of the Na$^{+}$ ions reduce the Fano effect because the lifetime of the phonon and the phonon-continuum interaction are decreased.
}
\kword{spin-Peierls transition, $\alpha'$-NaV$_2$O$_5$, Raman scattering, doping effect, Fano effect}
\def\runtitle{Raman scattering in the inorganic spin-Peierls system $\alpha '$-NaV$_{1-\delta}$V$_2$O$_5$}
\def\runauthor{H. {\sc Kuroe} {\it et al.}}
\begin{document}
\sloppy
\maketitle
\section{Introduction}
	$\alpha'$-NaV$_2$O$_5$ is one of the several phases in the Na$_x$V$_2$O$_5$ system.
Five bronze phases, 
$\alpha$ ($0 \leq x \leq 0.02$), 
$\beta$  ($0.02 \leq x \leq 0.40$),
$\alpha '$ ($0.70 \leq x \leq 1.00$),
$\eta$ ($1.28 \leq x \leq 1.45$) and 
$\chi$ ($1.68 \leq x \leq 1.82$), have been reported.\cite{Pouchard67}
	Since the $\beta$ phase shows the quasi-one-dimensional electrical conduction, it has been intensively investigated.\cite{Kanai82,Onoda82}
	Recently, the spin-Peierls (SP) transition at $T_{\rm SP}$ = 34 K was found in $\alpha'$-NaV$_2$O$_5$.\cite{Isobe96}
	This discovery followed that of CuGeO$_3$ in inorganic compounds and it has begun to attract new widespread attention.
	This phase transition occurs in a one-dimensional (1D) $S$ = 1/2 Heisenberg linear chain strongly interacting with the three-dimensional phonon system.\cite{Pytte74,Bary83}

	The structure of $\alpha '$-NaV$_2$O$_5$ is based on that of V$_2$O$_5$, which is built up from distorted trigonal bipyramidal coordination polyhedra of O$^{2-}$ ions around V$^{5+}$ ones.
	The VO$_5$ pyramids share edges to form (V$_2$O$_4$)$_{n}$ zigzag double chains along the $b$ direction and are closely linked along the $a$ direction through the shared corners, thus forming sheets in the $ab$ plane.\cite{Abello83}
	There are four pyramids in the unit cell of V$_2$O$_5$, and these are equivalent to each other.\cite{Backman61}
	The space group of V$_2$O$_5$ at room temperature belongs to $Pmmn$.
	In $\alpha '$-NaV$_2$O$_5$, there are two crystallographic different vanadium sites, which form two kinds of vanadium chains, respectively.
	The crystal structure has no inversion center and the space group belongs to $P2_1mn$ at room temperature.\cite{Carpy75}
	This compound is expected to be a quasi-1D spin system because the $S=1/2$ magnetic V$^{4+}$ chains are isolated from each other by the nonmagnetic V$^{5+}$ chains.

	In 1996, Isobe {\it et al.}\cite{Isobe96} reported that the magnetic susceptibility at high temperatures is well fitted to that of the typical 1D $S=1/2$ antiferromagnet with an exchange interaction $J= 560$ K and a $g$-factor $g \approx 2.0$.
	They found a decrease of the magnetic susceptibility below $T_{\rm SP} = 34$ K, thus suggesting the formation of a nonmagnetic spin-singlet ground state, and pointed out that the SP transition occurred at this temperature.
	Immediately after this report, the x-ray diffraction study in a single crystal showed that the superlattice reflections with a modulation vector ($1/2$, $1/2$, $1/4$) newly appeared at 10 K, indicating the formation of the $2a \times 2b \times 4c$ superlattice structure.\cite{Fujii96}
	The intensity of the superlattice reflection at (${3}/{2}$, ${1}/{2}$, ${11}/{4}$) was fitted to an equation of $(1-T/T_{\rm SP})^{2\beta }$, where $T_{\rm SP} = 35.27$ K and $\beta = 0.26 \pm 0.02$.
	The energy gap in the magnetic excitation spectrum was observed at about 9.8 meV with the wave vector $|{\bf Q}|$ = 1 \AA$^{-1}$ in a powder sample by means of inelastic neutron scattering.\cite{Fujii96,Yoshihama97}
	The SP transition in $\alpha '$-NaV$_2$O$_5$ was also observed by means of NMR,\cite{Ohama97b,Ohama97} ESR\cite{Vasilev97} and magnetic-resonance\cite{Schmidt97} measurements.
	The occurence of the SP transition in this system has been confirmed by these experiments.
	$T_{\rm SP}$ (= 34 K) of $\alpha '$-NaV$_2$O$_5$ is much higher than any other organic and inorganic SP systems.

	Raman scattering is a powerful tool to study the SP transition and it has been performed for an inorganic SP system CuGeO$_3$.\cite{Kuroe94b,Loosdreht96,Loa96,Lemmens96b,Muthukumar96,Kuroe97a}
	In the SP phase of CuGeO$_3$, two folded phonon modes at 368 and 818 cm$^{-1}$ were observed.
	In addition, the second-order magnetic Raman scattering appeared in the low-frequency Raman spectra.
	An asymmetric Raman peak at 30 cm$^{-1}$ was ascribed to the formation of the two-magnetic-excitation bound state.\cite{Kuroe94b,Sekine97,Sekine97b}
	In $\alpha '$-NaV$_2$O$_5$, eight new peaks were observed at 5 K by Raman scattering,\cite{Weiden97} and several new lines were observed below $T_{\rm SP}$ by infrared absorption,\cite{Popova97} indicating the formation of the superlattice.
	A broad Raman band was observed around 500 cm$^{-1}$ in the wide temperature range.
	Weiden {\it et al.} reported that it appeared in the $(b,b)$ geometry and assigned it to magnetic excitations.\cite{Weiden97}
	On the other hand, Golubchik {\it et al.} observed it in the $(a,a)$ geometry and assigned it to the $d$-$d$ electronic transition.
	They contradict each other, and it remains unresolved what the origin of the broad band is.

	Weiden {\it et al.} suggested an existence of the Fano resonance in the $(b,b)$ geometry Raman spectrum,\cite{Weiden97} which was due to the interaction between a phonon at about 530 cm$^{-1}$ and magnetic excitations around 500 cm$^{-1}$.
	The Fano effect takes place when a discreet phonon system interacts with a continuous electronic excitation, and it was observed in $p$-type Si\cite{Cerdeira73a,Cerdeira73b} and high-$T_{\rm c}$ superconductors.\cite{Thomsen88a,Cooper88,Cardona89,Thomsen88b}
	In the SP phase of CuGeO$_3$, Loa {\it et al}. stated that the origin of the asymmetric lineshape of the 105-cm$^{-1}$ peak is Fano resonance between the phonon and the magnetic excitation,\cite{Loa96} which is contrary to our assignment that it is ascribed to the van-Hove singularity in the spin-excitation spectrum.\cite{Kuroe94b}
	The origin of the 105-cm$^{-1}$ peak in CuGeO$_3$ is still controversial.
	In $\alpha '$-NaV$_2$O$_5$ system, Raman-scattering studies on the effect of the Na$^{+}$-ion deficiency on the Fano resonance are needed to clarify the origin of the continuous band.

	It is of interest to study the finite-length effect of the antiferromagnetic linear chain on the SP transition. 
	This effect has been extensively studied in CuGeO$_3$ by substituting Zn$^{2+}$ ions for Cu$^{2+}$ ones.\cite{Hase93} 
	In the Zn-doped CuGeO$_3$, $T_{\rm SP}$ and the lattice distortion of superlattice decreased at low concentration of Zn$^{2+}$ ions, and moreover a new phase with an antiferromagnetic order of the staggered spins appeared at low temperatures.\cite{Hase95,Lussier95,Oseroff95}
	Similar results were obtained in the Si-doped samples, where Si$^{4+}$ ions are substituted for the Ge$^{4+}$ ions and do not directly cut the one-dimensional linear spin chain.
	The suppression of the SP transition was observed by Raman scattering in the Zn- and Si-doped CuGeO$_3$. \cite{Kuroe96a,Kuroe96c,Lemmens97b,Weiden97b,Loosdrecht97,Fischer97}
	With increasing Zn or Si concentrations, the Raman intensity of the folded phonon mode decreased and its halfwidth increased.
	These reflect the decrease of the lattice distortion and the increase of the fluctuations of lattice dimerization, respectively.\cite{Kuroe96a,Kuroe96c}
	The SP-gap mode became observable in the first-order Raman process because of the breakdown of the selection rule, and the energy of the SP gap decreased with increasing dopant concentrations.\cite{Sekine97,Sekine97b}

	In $\alpha '$-NaV$_2$O$_5$, on the other hand, the vanadium ion in the vicinity of the defect of the Na$^{+}$ ion changes to a nonmagnetic V$^{5+}$.
	This directly cuts the V$^{4+}$-ion magnetic linear chain.
	$T_{\rm SP}$ in the sample with defects of 1-\% Na$^{+}$ ions was determined to be about 32 K by means of the magnetic susceptibility, while no indication of the SP transition, i.e., no decrease of the magnetic susceptibility, was detected in the sample with defects of 10-\% Na$^{+}$ ions.\cite{Isobe97}
	The low-temperature antiferromagnetic phase, which was observed in Zn- and Si-doped CuGeO$_3$, has never been found in $\alpha '$-Na$_{1-\delta}$V$_2$O$_5$.
	Very recently, the temperature dependence of the superlattice intensity was determined by means of x-ray diffraction with the single crystals of $\alpha '$-Na$_{1-\delta}$V$_2$O$_5$, where $0 < \delta < 0.05$.\cite{Nakao97}
	The results clearly show that $T_{\rm SP}$ and the superlattice intensity, which is approximately proportional to the square of the amplitude of the lattice distortion, decreased with increasing $\delta$.
	These results are qualitatively consistent with those in the impurity-doped CuGeO$_3$.
	It was also observed that the intensity of the critical scattering around $T_{\rm SP}$ became strong when the Na$^{+}$ ions defect from the crystal, suggesting the increment of the fluctuations of lattice.

	In our preliminary report, we showed Raman spectra of $\alpha '$-Na$_{1-\delta}$V$_2$O$_5$ with various temperatures and $\delta$.\cite{Kuroe96c}
	In this paper, we measure Raman scattering in the single crystals of $\alpha '$-Na$_{1-\delta} $V$_2$O$_5$ with $\delta$= 0, 0.01 and 0.1 in order to clarify the origin of the broad band around 500 cm$^{-1}$ and to study the effects of the Na$^+$-ion deficiency on the SP transition and the Fano effect.
	Moreover we discuss the origin of new Raman peaks in the SP phase of $\alpha '$-NaV$_2$O$_5$.
	Taking into account the results of the polarization characteristics of Raman spectra in the uniform and the SP phases, we propose possible symmetries of the lattice in the SP phase.

\section{Experimental Details}
	The detail of the sample preparation was described in Ref. \ref{Isobe97b}.
	The size of the single crystals of $\alpha '$-Na$_{1-\delta}$V$_2$O$_5$ are $2.0 \times 4.7 \times 0.2$, $1.8 \times 6.3 \times 0.5$ and $1.8 \times 4.0 \times 0.8$ mm$^{3}$ for $\delta$ = 0, 0.01 and 0.1 along the $a$, $b$ and $c$ axes, respectively. 
	The value of $\delta$ is the nominal concentration, and it was roughly checked by the magnetic susceptibility with the sample taken from the same lot.
	The $ab$ face of each sample is attached by dilute varnish on the sample holder in a closed-cycle cryostat.
	With this cryostat, we measured Raman spectra between 15 and 300 K.
	The temperature of the sample is controlled by a PID temperature controller.
	The additional thermometer is attached on the back of the sample holder, and the temperature of the sample is equal to that of the sample holder within $\pm$0.1 K.

	We used the 5145-{\AA} line of the Ar$^{+}$-ion laser as an incident light.
	We passed the laser light through two filters in order to eliminate plasma lines of the laser.
	We used the quasibackscattering geometry because the single crystals are opaque.
	The laser line was incident onto the $ab$ face of the sample in the cryostat.
	Since the incident and scattered lights went approximately along the $c$ axis within the sample, we observed transversed optical (TO) phonon modes above $T_{\rm SP}$.\cite{Golubchik97}
	In polarization measurements, we set a polarizer in front of the sample and analyzed the scattered light by a prism polarizer at the entrance of Jobin-Yvon U1000 double-grating monochromator.
	When the incident and the scattered photons are respectively polarized along $x$ and $y$ axes of the crystal, where $x$ and $y$ respectively correspond to the $a$ or $b$ axes in the uniform phase, we represent the geometry of this experiment as $(x,y)$.
	The scattered light is detected by a photon counting system which is controlled with a micro-computer.
	The spectra shown in this study were obtained as the results of averages at least over three runs, because the scattered light is very weak.

\section{Results}
\begin{figure}
\begin{center}
\epsfile{file=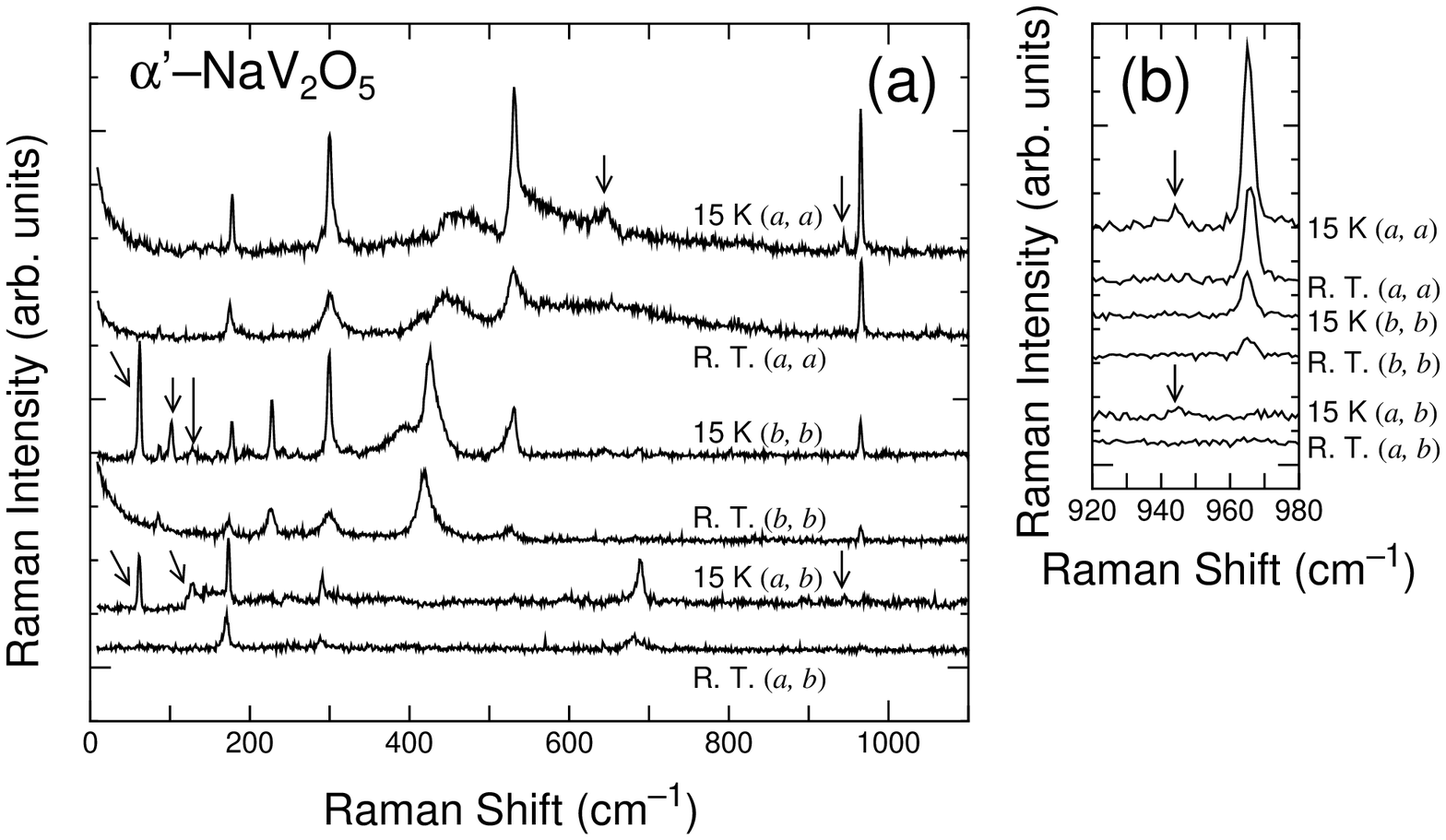,width=0.9\textwidth}
\end{center}
\caption{Polarization characteristics of Raman spectra in $\alpha ' $-NaV$_2$O$_5$ at room temperature and 15 K (a) and the expanded Raman spectra between 920 and 980 cm$^{-1}$ (b).}
\label{rt}
\end{figure}
\begin{fulltable}
\begin{fulltabular}{*{5}{c}}
\hline
$(a,a)$ [cm$^{-1}$] & $(b,b)$ [cm$^{-1}$] & $(a,b)$ [cm$^{-1}$] & origin & symmetry at R. T. \\
\hline
--- &  *62 &  *62 & 1st order SP gap & --- \\
87  &   87 &  --- & phonon & $A_{1}$ \\
--- & *102 &  --- & folded phonon & --- \\
--- & *128 & *128 & 2nd order SP gap & --- \\
--- &  --- &  170 & phonon & $A_{2}$ \\
174 &  174 &  --- & phonon & $A_{1}$ \\
--- &  228 &  --- & phonon & $A_{1}$ \\
--- &  --- &  290 & phonon & $A_{2}$ \\
300 &  300 &  --- & phonon & $A_{1}$ \\
--- &  {}*393 &  --- & ? & --- \\
--- &  418(R.T)                 & --- & ? & --- \\
    &  $\rightarrow$ 425 (15 K) &     &   & \\
444 &  --- &  --- & electronic continuum & --- \\
531 &  531 &  --- & phonon & $A_{1}$ \\
{}*646 & --- &  --- & folded phonon & --- \\
--- &  --- &  680 & phonon & $A_{2}$ \\
{}*944 & --- & {}*944 & folded phonon & --- \\
965 &  965 &  --- & phonon & $A_{1}$ \\
\hline
\end{fulltabular}
\caption{Peak positions in the $(a,a)$, $(b,b)$ and $(a,b)$ geometries at room temperature and 15 K. The peaks observed only at 15 K are marked by the asterisk(*).}
\label{spectra}
\end{fulltable}
	Figure \ref{rt}(a) shows polarized Raman spectra of $\alpha ' $-NaV$_2$O$_5$ between 15 and 1100 cm$^{-1}$ at room temperature and 15 K.
	The Raman spectra between 920 and 980 cm$^{-1}$ are expanded into Fig. \ref{rt}(b).
	We observed five new peaks at 62, 102, 128, 646 and 944 cm$^{-1}$ at 15 K, which were indicated by arrows in Figs. \ref{rt}(a) and \ref{rt}(b).
	The frequencies of the observed peaks are listed in the Table \ref{spectra}.

	In the $(a,a)$ geometry, we observed four sharp Raman peaks at 87, 174, 300 and 965 cm$^{-1}$ at room temperature.
	In addition, the broad Raman band, on which two asymmetric peaks at 444 and 531 cm$^{-1}$ were superimposed, was also observed.
	At 15 K, we observed two new peaks at 646 and 944 cm$^{-1}$.

	In the $(b,b)$ geometry, we observed five sharp and symmetric peaks at 87, 174, 228, 300 and 965 cm$^{-1}$ at room temperature.
	At 15 K, the new Raman peaks appeared at 62, 102, 128 and 228 cm$^{-1}$.
	The 418-cm$^{-1}$ peak at room temperature in the $(b,b)$ geometry shifted to 425 cm$^{-1}$ at 15 K, and a small shoulder appeared at 15 K.
	The halfwidth of this peak is much broader than the phonon mode at 300 cm$^{-1}$ at 15 K.

	In the $(a,b)$ geometry, we observed three peaks at 170, 290 and 680 cm$^{-1}$ at room temperature.
	At 15 K, new Raman peaks appeared at 62, 102, 128 and 944 cm$^{-1}$.
	Moreover we observed a continuous band between 130 and 400 cm$^{-1}$.
	It consists of a rather strong band below about 200 cm$^{-1}$ and a veryweak band between 200 and 400 cm$^{-1}$.
	As shown in Fig. \ref{rt}(b), the 944-cm$^{-1}$ peak is observed in both the $(a,a)$ and $(a,b)$ geometries.

\begin{fullfigure}
\begin{center}
\epsfile{file=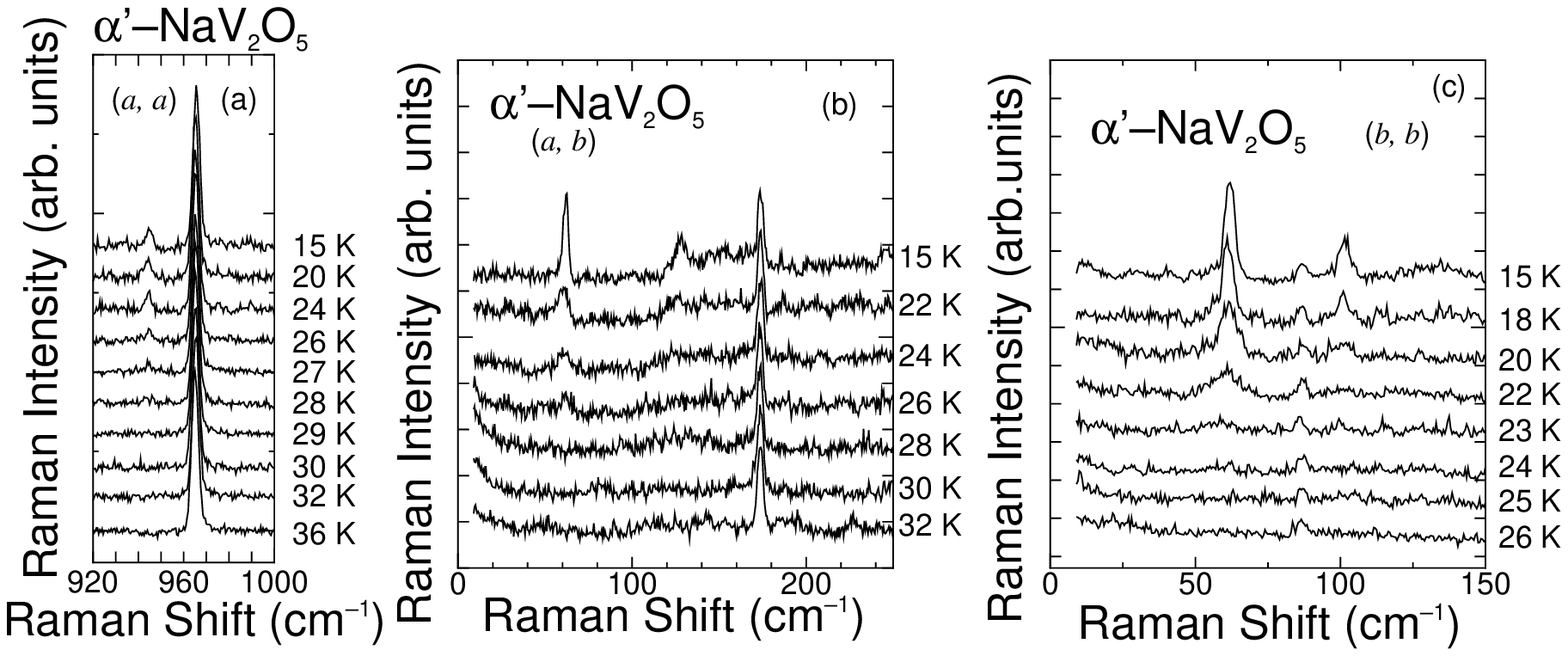,width=0.9\textwidth}
\end{center}
\caption{Detailed temperature dependence of Raman spectra in the $(a,a)$ geometry (a), the $(a,b)$ one (b) and the $(b,b)$ one (c).}
\label{tdependence}
\end{fullfigure}
	Figures \ref{tdependence}(a), \ref{tdependence}(b) and \ref{tdependence}(c) show the detailed temperature dependence of the new peaks in the $(a,a)$, $(a,b)$ and $(b,b)$ geometries, respectively.
	The 62- and 128-cm$^{-1}$ peaks were observed in both the $(a,b)$ and $(b,b)$ geometries.
	The 62-cm$^{-1}$ peak seems to have a tail on the lower-energy side, and its peak position slightly shifts to the lower-energy side with increasing temperature.
	This asymmetric Raman spectra probably consists of two Raman peaks of which frequencies are very close to each other.
	The new peaks in the $(a,a)$, $(a,b)$ and $(b,b)$ geometries disappeared above 28, 28 and 25 K, respectively.
	All of the Raman spectra were measured with the same power of the incident laser light.
\begin{figure}
\begin{center}
\epsfile{file=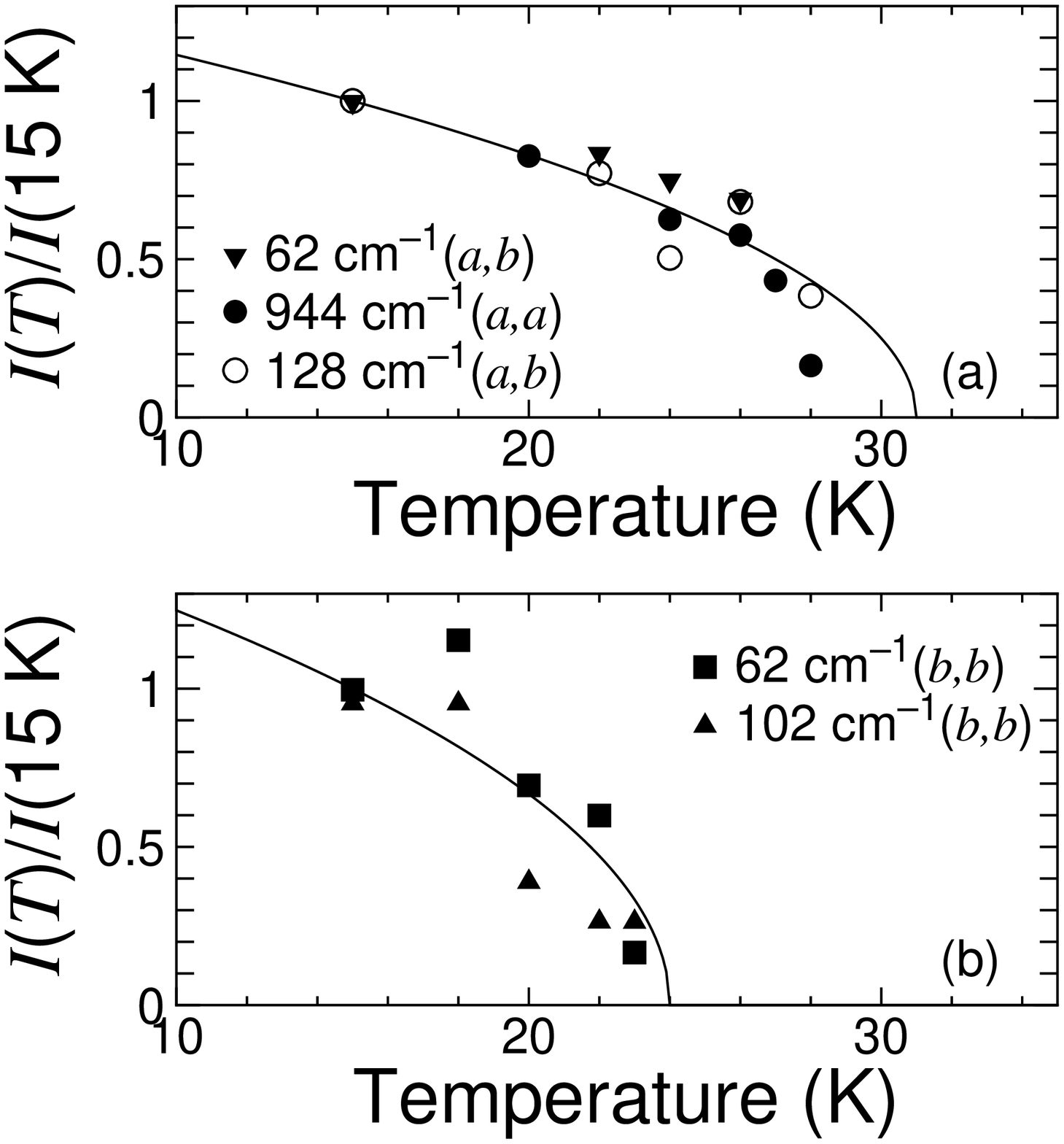,width=0.5\textwidth}
\end{center}
\caption{
Raman intensities $I$ of the peaks 
at 62 cm$^{-1}$ in the $(a,b)$ geometry (closed reversed triangles) and 
at 994 cm$^{-1}$ in the $(a,a)$ geometry (closed circles) 
and the Raman band in the $(a,b)$ geometry (open circles) (a), 
and $I$ of the peaks 
at 62 cm$^{-1}$ in the $(b,b)$ geometry (closed squares) and 
at 102 cm$^{-1}$ in the $(b,b)$ geometry (closed triangles) (b).
}
\label{Rintensity}
\end{figure}

	Figures \ref{Rintensity}(a) and \ref{Rintensity}(b) show the temperature dependence of the integrated intensities of the peaks observed in the $(a,a)$ and $(a,b)$ geometries and that in the $(b,b)$ one, respectively.
	All the intensities are normalized by the value at 15 K.
	Here we assumed the population factor to be unity below 30 K.
	Because it is difficult to separate the 128 cm$^{-1}$ peak in the $(a,b)$ geometry from the continuous band lying just above it, especially around $T_{\rm SP}$, we show the intensity of the broad band below 200 cm$^{-1}$, instead of the intensity of the 128 cm$^{-1}$ peak.
	The temperature at which the new peaks in the $(a,a)$ and $(a,b)$ geometries disappear (28 K) is lower than $T_{\rm SP}$ obtained from the magnetic susceptibility\cite{Isobe96} and the x-ray diffraction.\cite{Fujii96}
	But the temperature dependence of the Raman intensity is well described by a function of $\sqrt{T_{\rm SP}-T}$, when we assume $T_{\rm SP}$ to be 31 K.
	Since the new peaks weakened around $T_{\rm SP}$, as shown in Figs. \ref{tdependence}(a) and \ref{tdependence}(b), these peaks smeared out above 28 K.

	As shown in Fig. \ref{tdependence}(c), the peaks in the $(b,b)$ geometry was not observed above 25 K, which is much lower than $T_{\rm SP}$.
	The difference probably comes from the local temperature rise due to the incident laser line because the single crystal of $\alpha '$-NaV$_2$O$_5$ is opaque.
	When the incident light is polarized along the $b$ axis, the light may be incident approximately with the Brewster angle in the quasibackscattering geometry.
	On such a condition, almost all of the incident light can penetrate into the sample, and it may cause the local temperature rise.
	It is difficult to measure Raman scattering with weaker incident light because the Raman spectra of $\alpha '$-NaV$_2$O$_5$ have weak intensities.
	Consequently, the new peaks at 62, 102, 128, 646 and 944 cm$^{-1}$ in the SP phase are closely correlated with the SP transition.
	The Raman peaks at 102, 646 and 944 cm$^{-1}$ are assigned to the folded-phonon modes, and, as will be discussed later, the peaks at 62 and 128 cm$^{-1}$ to the magnetic excitations. 
\begin{figure}
\begin{center}
\epsfile{file=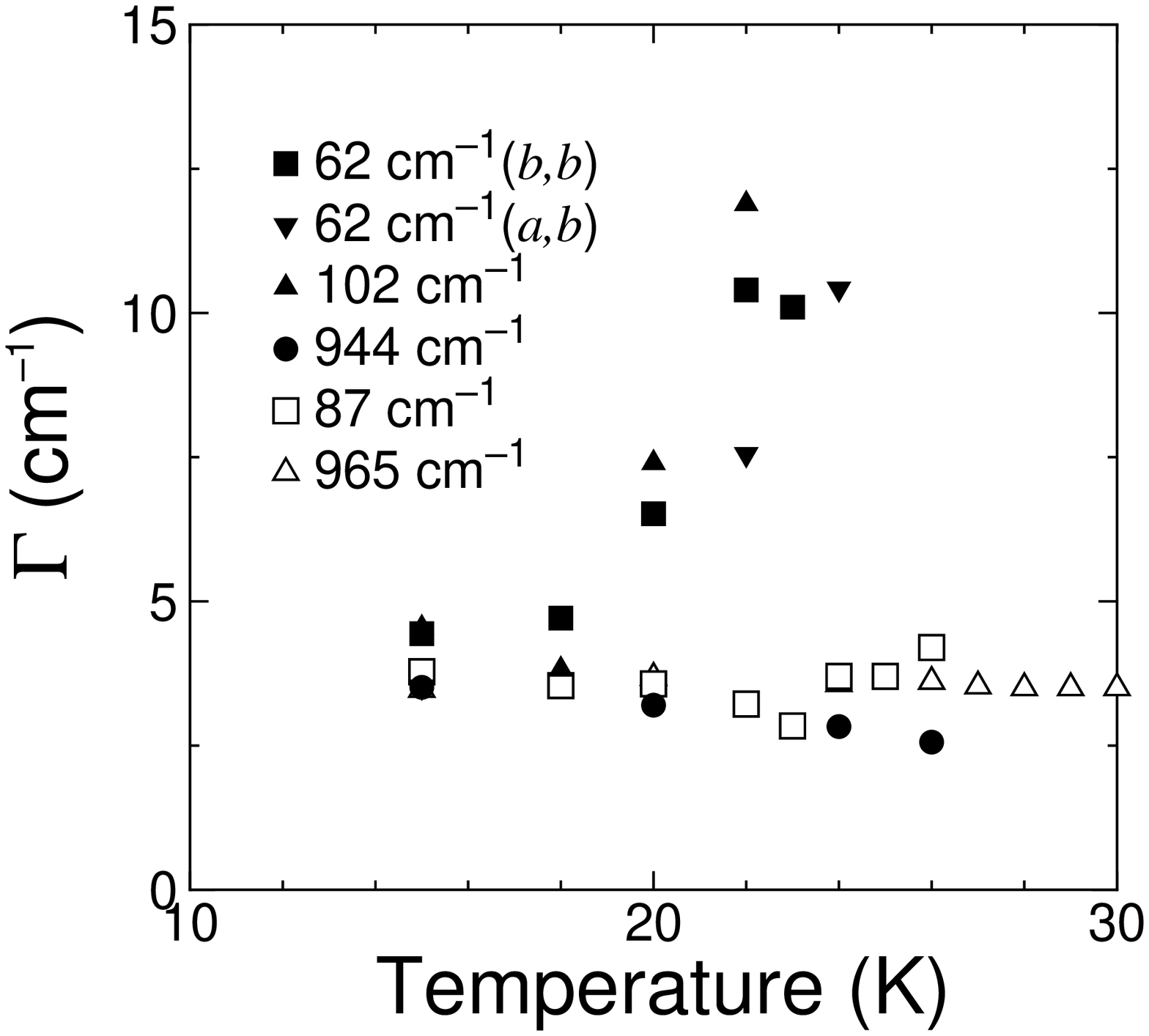,width=0.5\textwidth}
\end{center}
\caption{Full widths $\Gamma$ at half maxima of the peaks at 62 cm$^{-1}$ in the $(a,b)$ (closed reversed triangles) and in the $(b,b)$ one (closed squares), 87 cm$^{-1}$ (open squares), 102 cm$^{-1}$ (closed triangles), 944 cm$^{-1}$ (closed circles) and 964 cm$^{-1}$ (open triangles).}
\label{Ghalfwidth}
\end{figure}

	Figure \ref{Ghalfwidth} shows the full widths at half maxima (FWHM) of several Raman peaks.
	The FWHMs of the 102- and the 646-cm$^{-1}$ peaks observed only in the SP phase increase as the temperature is increased toward $T_{\rm SP}$.	
	In general, the lifetime of the folded-phonon mode shortens around $T_{\rm SP}$ because of the large fluctuations.
	The temperature dependence of the FWHM of the 944-cm$^{-1}$ peak is similar to that of the Raman-active $A_{1}$ phonons at 87 and 965 cm$^{-1}$.
	Since the frequency of the 944-cm$^{-1}$ peak is much higher than those of other folded phonon modes, the effects of fluctuations on the 944-cm$^{-1}$ peak may be weaker than those in any other peaks.
\begin{figure}
\begin{center}
\epsfile{file=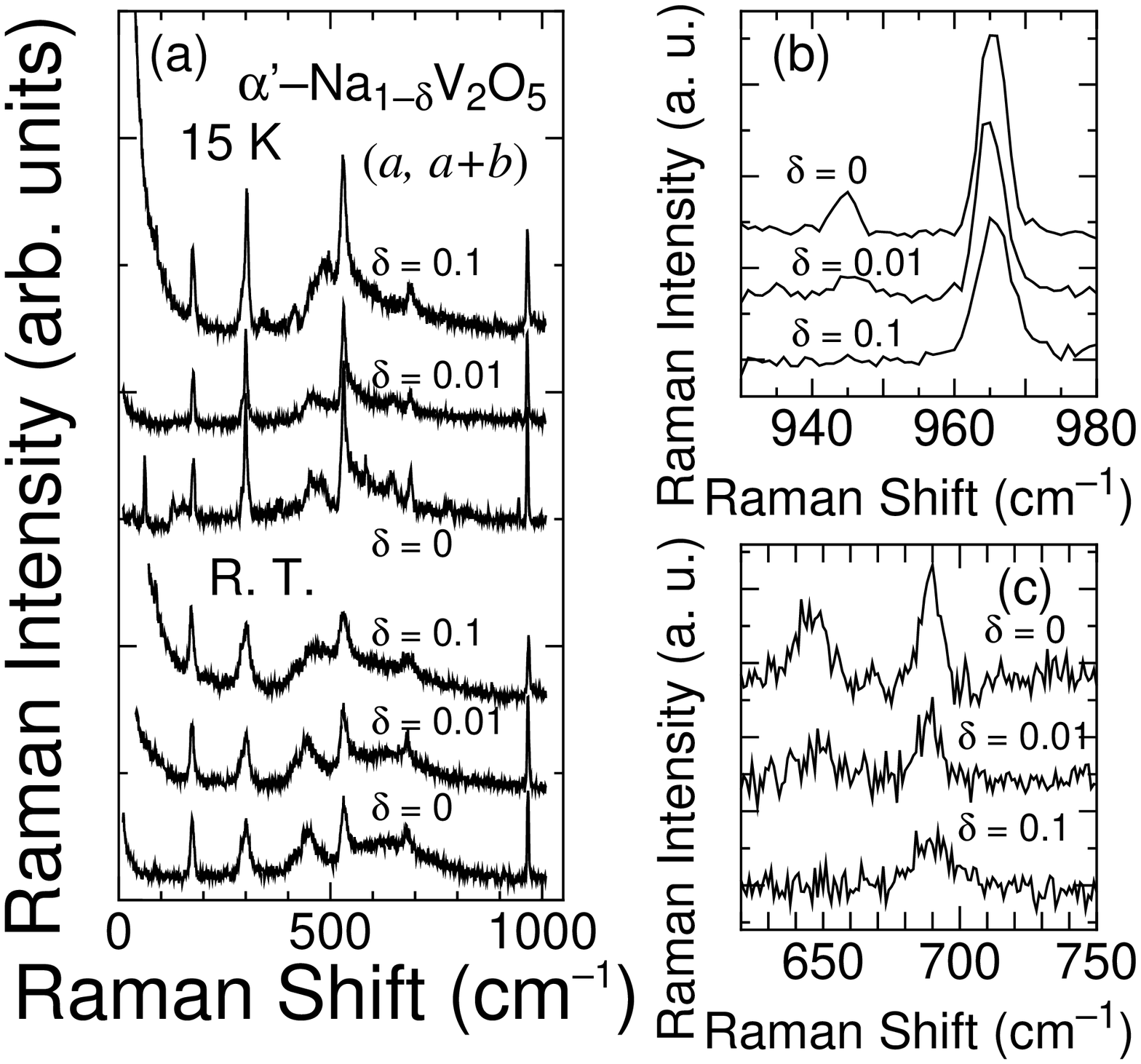,width=0.5\textwidth}
\end{center}
\caption{
Raman spectra in $\alpha '$-Na$_{1-\delta}$V$_2$O$_5$ ($\delta$ = 0, 0.01 and 0.1) at room temperature and 15 K (a) and the normalized Raman spectra between 930 and 980 cm$^{-1}$ (b) and between 630 and 750 cm$^{-1}$ (c).
}
\label{defectedsamples}
\end{figure}

	To discuss the effects of the nonmagnetic impurity doping on the folded-phonon mode, we measured Raman scattering of $\alpha '$-Na$_{1-\delta}$V$_2$O$_5$.
	Figure \ref{defectedsamples}(a) shows Raman spectra in $\alpha '$-Na$_{1-\delta}$V$_2$O$_5$ ($\delta$ = 0, 0.01 and 0.1) at room temperature and 15 K in the $(a, a+b)$ geometry.
	The five new Raman peaks and the continuous band between 130 and 400 cm$^{-1}$ weakened in the sample of $\delta$ = 0.01 and these were not observed in the sample of $\delta$ =  0.1. 
	On the other hand, the intensity of the tail of the direct scattering by the incident laser line in the low-frequency Raman spectra at room temperature and 15 K rapidly increases as the defects of Na$^{+}$ ions are increased.
	The quasielastic scattering due to the fluctuations of the energy density in the spin system was observed in CuGeO$_3$\cite{Kuroe97a} and it is one of the possible origins for the strong tail in $\alpha '$-NaV$_2$O$_5$.
	For another possibility, single particle excitations of carriers, which were observed in $n$-GaAs,\cite{Pinczuk71} give such a strong tail of the incident laser.
	Moreover the asymmetric peak at 531 cm$^{-1}$ becomes almost symmetric with increasing Na$^{+}$ defect.

\begin{fulltable}
\begin{fulltabular}{*{6}{c}}
$\delta$ & $\Gamma (944 {\rm cm}^{-1})$ & $\Gamma (965 {\rm cm}^{-1})$ & $I (944 {\rm cm}^{-1})$ & $I (965 {\rm cm}^{-1})$ & $I (944 {\rm cm}^{-1})$ / $I (965 {\rm cm}^{-1})$ \\
\hline
0    & 4.2  & 3.5  & 361   & 2100 & 0.18  \\
0.01 & 5.4  & 4.8  & 234   & 1800 & 0.13  \\
0.1  &  --- & 6.3  &     0 & 2500 &     0 \\
\hline
\hline
$\delta$ & $\Gamma (646 {\rm cm}^{-1})$ & $\Gamma (690 {\rm cm}^{-1})$ & $I (646 {\rm cm}^{-1})$ & $I (690 {\rm cm}^{-1})$ & $I (646 {\rm cm}^{-1})$ / $I (690 {\rm cm}^{-1})$ \\
\hline
0    & 12   & 8.5  & 980   & 880   & 1.1   \\
0.01 & 16   & 10   & 620   & 740   & 0.84  \\
0.1  &  --- & 14   &     0 & 1400  &     0 \\
\end{fulltabular}
\caption{Full widths at half maxima $\Gamma$ and Raman intensities $I$ of the folded phonon modes at 646 and 944 cm$^{-1}$ and the Raman-active phonon modes at 690 and 965 cm$^{-1}$ in samples of $\delta$ = 0, 0.01 and 0.1.}
\label{defectfoldedphonon}
\end{fulltable}
	Figures \ref{defectedsamples}(b) and \ref{defectedsamples}(c) show Raman spectra of the 646- and the 944-cm$^{-1}$ folded-phonon modes at 15 K of which intensities are normalized by the phonons at 680 and 965 cm$^{-1}$, respectively, in order to discuss the effects of the Na$^{+}$-ion deficiency on the Raman intensity quantitatively.
	Here the background, which contains the dark count of the photomultiplier and the effects of the continuous band, is assumed to be a function of $a \omega^2 + b\omega +c$.
	The FWHM $\Gamma$ and the Raman intensity $I$ are listed in Table \ref{defectfoldedphonon} and show clearly that the folded phonon mode weakened and broadened in the sample of $\delta = 0.01$.
	The weakening and broadening of the folded phonon modes were also observed in the folded phonon at 368 cm$^{-1}$ in CuGeO$_3$.\cite{Kuroe96a,Kuroe96c}
	The increase of the FWHMs with increasing $\delta$, which indicates the enhancement of the lattice fluctuations, is consistent with the increase of the critical scattering obtained by x-ray diffraction.\cite{Nakao97}
	We did not observe the folded-phonon modes at 646 and 944 cm$^{-1}$ in the sample of $\delta = 0.1$, indicating that the SP transition does not occur in this sample above 15 K.
	This result is consistent with those of the magnetic susceptibility\cite{Isobe97} and x-ray diffraction.\cite{Nakao97}
\begin{figure}
\begin{center}
\epsfile{file=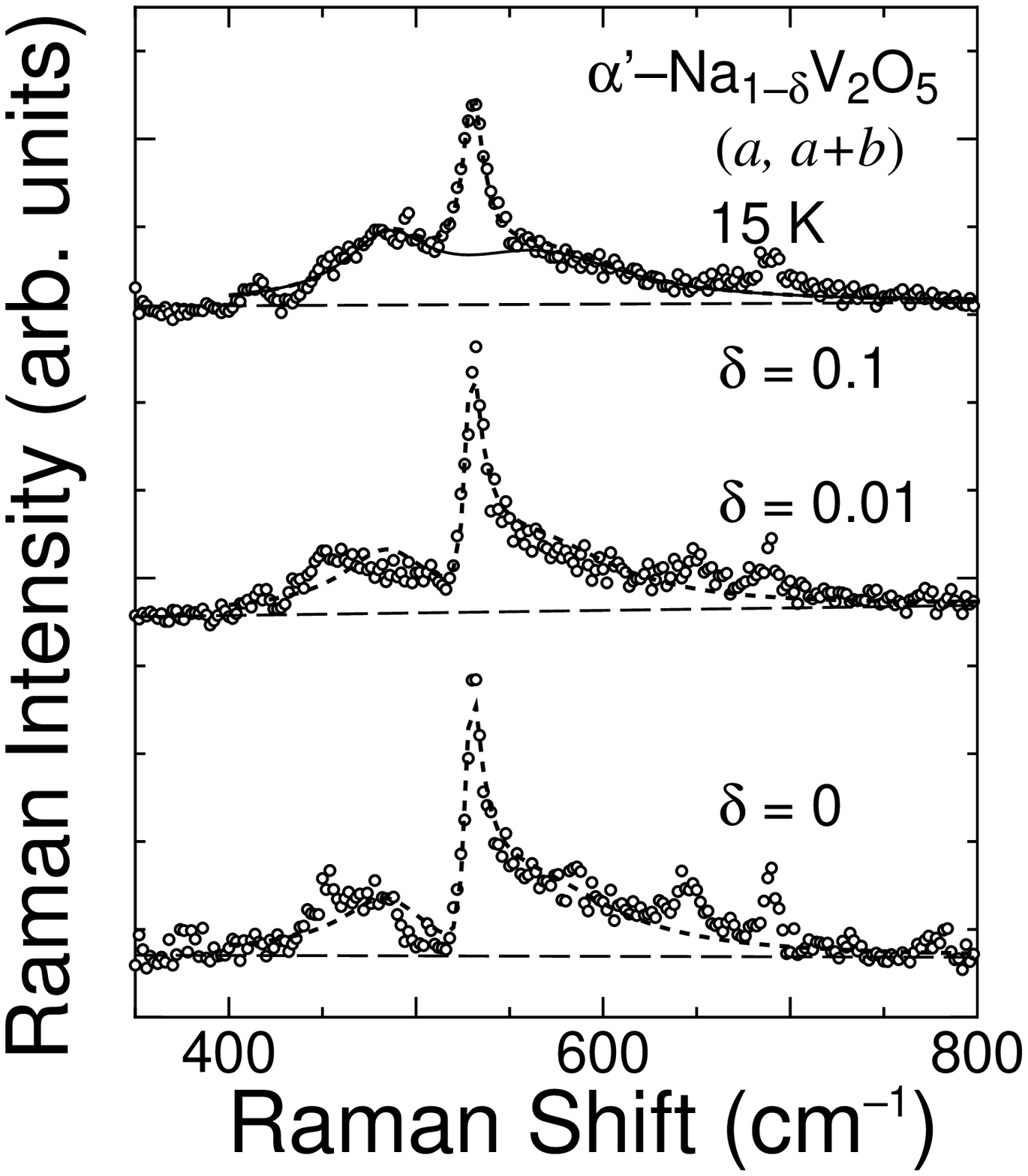,width=0.5\textwidth}
\end{center}
\caption{
Fano asymmetry of Raman spectra in $\alpha '$-Na$_{1-\delta}$V$_2$O$_5$ ($\delta$ = 0, 0.01 and 0.1).
The solid curve, the dotted curves, and the dashed lines denote the density of state of the electronic excitation, the calculated Raman spectra, and the background, respectively.
}
\label{fanoeffect}
\end{figure}

	Figure \ref{fanoeffect} shows Raman spectra of the continuous band between 350 and 800 cm$^{-1}$ in the samples of $\delta$ = 0, 0.01 and 0.1.
	The Raman intensity was normalized by that of the $A_1$ phonon mode at 680 cm$^{-1}$.
	We observed the broad band around 500 cm$^{-1}$, the $A_{1}$ phonon peak at 531 cm$^{-1}$, the folded phonon mode at 646 cm$^{-1}$ and the $A_2$ phonon mode at 680 cm$^{-1}$ in the sample of $\delta$ = 0 and 0.01.
	In the sample of $\delta = 0.1$, the folded phonon mode was not observed and the Raman peak at 531 cm$^{-1}$ is almost symmetric in lineshape.
	The small asymmetric peak at 455 cm$^{-1}$ appears in the sample of $\delta$ = 0 and 0.01.
	On the other hand, a dip at 431 cm$^{-1}$ appears in all the samples.
	The dashed lines in Fig. \ref{fanoeffect} denote the background coming from the dark count of the photomultiplier and higher-order Raman scattering.
	The dotted curves and the solid one in Fig. \ref{fanoeffect} will be described later.
\begin{figure}
\begin{center}
\epsfile{file=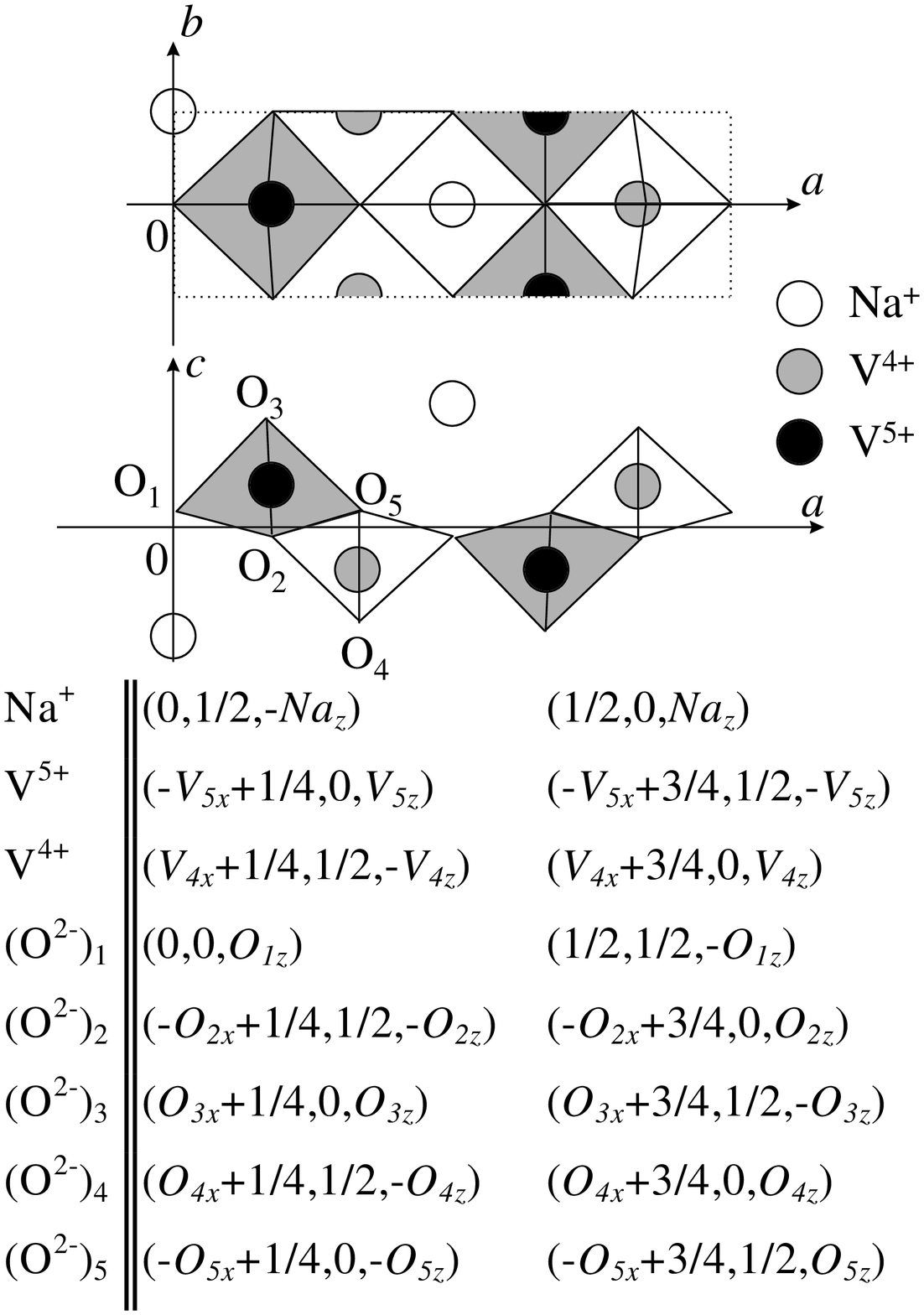,width=0.5\textwidth}
\end{center}
\caption{
Schematic lattice structure of $\alpha '$-NaV$_2$O$_5$ at room temperature.
The positions of the ions are also listed.
}
\label{unitcell}
\end{figure}

\section{Discussion}
	Figure \ref{unitcell} shows the unit cell of $\alpha '$-NaV$_2$O$_5$ at room temperature and the positions of ions in the lattice.
	One can easily see that all the positions of ions can be written as a form of $(A+1/4,0,B)$ or $(A+3/4,1/2,-B)$.
	The symmetry of the lattice at room temperature is $C_{2v}^{7}(P2_{1}mn)$.
	The $2_{1}$ screw operator along the $a$ axis, the mirror in the $ac$ plane and the glide operator in the $ab$ plane transfer a point $(x,y,z)$ to $(x+1/2,-y+1/2,-z)$, $(x,-y,z)$ and $(x+1/2,y+1/2,-z)$, respectively.
	
	Since the space group of $\alpha ' $-NaV$_2$O$_5$ at room temperature is $C_{2v}^{7}(P2_1mn)$, the following long-wavelength optical modes are expected from the factor group analysis;
\begin{equation}
\Gamma_{\rm opt} = 15 A_1 + 7 A_2 + 15 B_1 + 8 B_2\ .
\end{equation}
All the phonon modes are Raman active and their Raman tensors are written as
\[
A_1  : \left(
	\begin{array}{ccc}
	a &  &  \\
	  & b&  \\
	  &  & c\\
	\end{array}
	\right) \ ,\ 
A_2  : \left(
	\begin{array}{ccc}
	  & d&  \\
	 d&  &  \\
	  &  &  \\
	\end{array}
	\right) \ ,
\]
\[
B_1  : \left(
	\begin{array}{ccc}
	  &  & e\\
	  &  &  \\
	 e&  &  \\
	\end{array}
	\right) \ ,\ 
B_2  : \left(
	\begin{array}{ccc}
	  &  &  \\
	  &  & f\\
	  & f&  \\
	\end{array}
	\right) \ .
\]
	In the present experiments, the $A_1$ modes appear strongly in the $(a,a)$ and $(b,b)$ geometries and the $A_{2}$ modes appear in the $(a,b)$ one.
	As summarized in Table \ref{spectra}, we assign the peaks at 87, 174, 228, 300, 531 and 965 cm$^{-1}$ and those at 170, 290 and 680 cm$^{-1}$ to $A_1$ and $A_2$ phonon modes, respectively.
	These peaks corresponded to the TO $A_1$ phonon modes at 89, 175, 231, 300, 531 and 970 cm$^{-1}$ and the TO $B_1$ phonon modes at 172, 289 and 678 cm$^{-1}$ observed by Golubchik {\it et al.}, respectively.\cite{Golubchik97}
	We did not observe the Raman peaks at 215 and 254 cm$^{-1}$ which were reported by Golubchik {\it et al.} probably because of their weak intensities.
	Golubchik {\it et al.} reported additional TO $A_1$ phonon modes at 415, 421, 435 and 449 cm$^{-1}$ at room temperature.
	But in our spectrum, these did not split and correspond to a broad peak at 444 cm$^{-1}$ with a dip at 423 cm$^{-1}$.
	The halfwidth of the 425-cm$^{-1}$ peak at 15 K in the $(b,b)$ geometry seems to be broader than any other phonon peaks.
	The shoulder appeared at low temperatures, and the frequency shift (= 6 cm$^{-1}$) is considerably large between 15 K and room temperature. We think that it is not a usual phonon mode.
	Our result of the peak frequencies agrees with that of Weiden {\it et al.},\cite{Weiden97} but that of the polarization characteristics contradicts theirs, and the difference clearly comes from the quality of the samples, because their Raman spectra in the $(b,b)$ geometry contains the components of other geometries.

	Let us discuss the symmetry of the lattice structure in the SP phase.
	We observed the peak at 944 cm$^{-1}$ below $T_{\rm SP}$ in the $(a,a)$ and $(a,b)$ geometries, and it is assigned to the folded phonon.
	This result indicates that the symmetry of $\alpha '$-NaV$_2$O$_5$ in the SP phase should be triclinic or monoclinic because the phonon Raman tensor of the 944-cm$^{-1}$ peak should have diagonal and offdiagonal components simultaneously.
	The magnetic Raman peak at 62 cm$^{-1}$ is also observed in the $(a,a)$ and $(a,b)$ geometries, supporting the triclinic or monoclinic structure in the SP phase.
	To realize the monoclinic symmetry in the SP phase, two of these symmetry operations should be broken, i.e., the possible symmetries in the SP phase are $C_{2}^{2}(P2_{1})$, $C_{s}^{1}(Pm)$ and $C_{s}^{2}(Pn)$.
	When all the symmetries are broken in the SP phase, the lattice symmetry becomes triclinic $C_{1}^{1}(P1)$.
\begin{figure}
\begin{center}
\epsfile{file=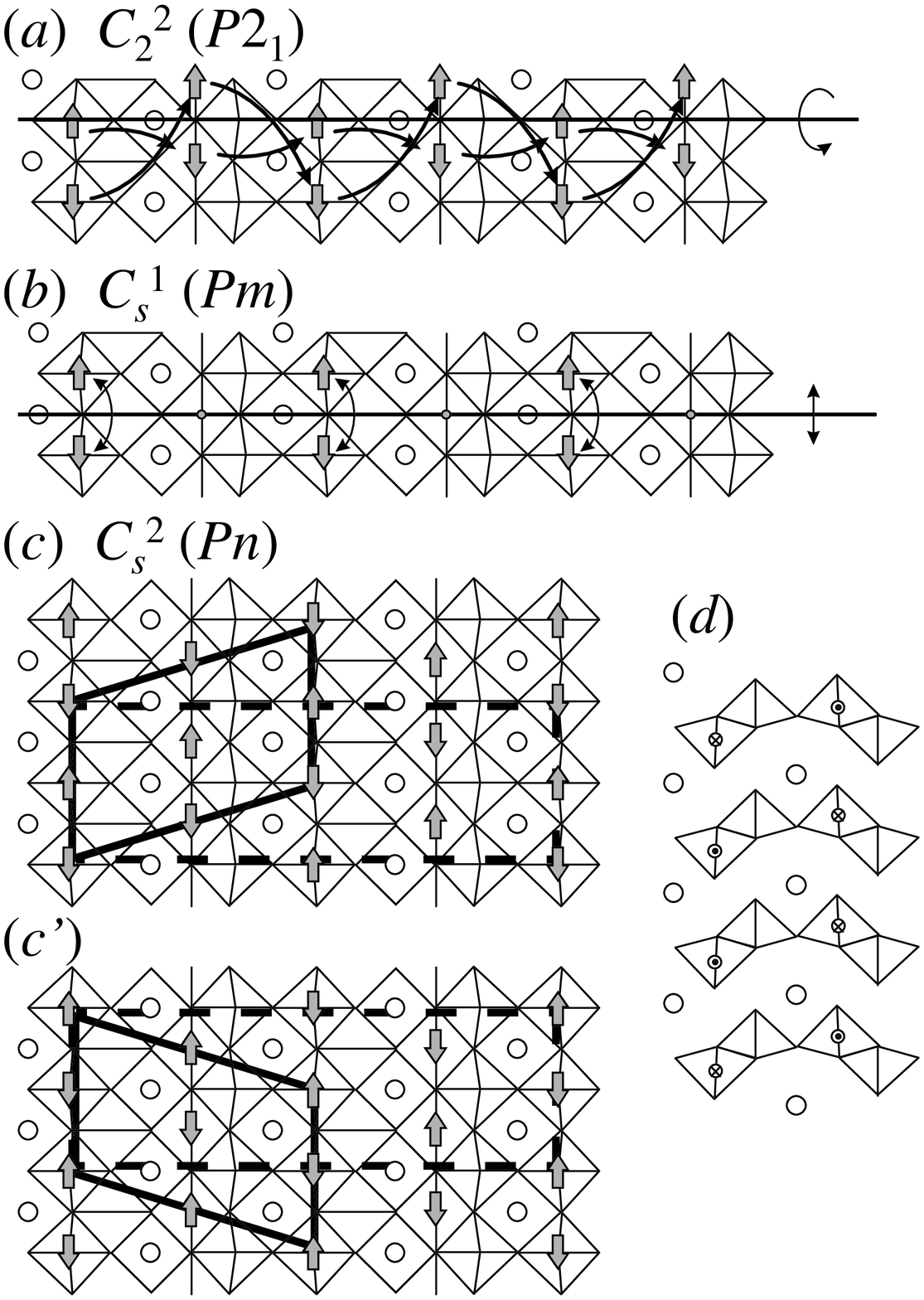,width=0.5\textwidth}
\end{center}
\caption{
Lattice distortions of the V$^{4+}$ ions along the $b$ axis when the symmetry of the lattice is assumed to be $C_{2}^{2}$ (a), $C_{s}^{1}$ (b) and $C_{s}^{2}$ (c and c$'$).
The structure in the $C_{s}^{2}$ symmetry projected on the $ac$ plane is shown in (d).
}
\label{latdisplacement}
\end{figure}

	Figures \ref{latdisplacement}(a), \ref{latdisplacement}(b), \ref{latdisplacement}(c) and \ref{latdisplacement}(c$'$) show the possible displacements of V$^{4+}$ ions which dimerize the lattice along the $b$ axis.
	For simplify, we neglect the displacement of the V$^{4+}$ ions perpendicular to the magnetic chain and that of other ions (V$^{5+}$, Na$^{+}$ and O$^{2-}$).
	This assumption is reasonable because the displacements of the V$^{4+}$ ions in the $ac$ plane do not change the distance between the V$^{4+}$ ions.
	For another possibility, one can infer that the O$^{2-}$ ion between the V$^{4+}$ ions shifts and it changes the path of the superexchange interaction.
	Our results for the V$^{4+}$ ions can be easily expanded into the displacement of O$^{2-}$ ions between the nearest neighbor V$^{4+}$ ions in the magnetic chain.

	As shown in Fig. \ref{latdisplacement}(a), the $2a \times 2b \times 4c$ superstructure could not be formed when the symmetry in the SP phase is $C_{2}^{2}$.
	When the $a$ axis is doubled by the displacements of another ions, the symmetry of the lattice becomes $C_{1}^{1}$, because the mirror and the glide plane have been already broken.
	Consequently, the symmetry in the SP phase is not $C_{2}^{2}$.

	As shown in Fig. \ref{latdisplacement}(b), half of the V$^{4+}$ ions in the lattice do not move when the symmetry in the SP phase is $C_{s}^{1}$, because these ions are located on the mirror plane.
	Since this result contradicts that of the magnetic susceptibility that all of spins die at 0 K,\cite{Isobe96} we think that the lattice in the SP phase does not belong to $C_{s}^{1}$.

	Figures \ref{latdisplacement}(c) and \ref{latdisplacement}(c$'$) show two possible displacements of the lattice, assuming the $C_{s}^{2}$ symmetry in the SP phase.
	In both the cases, the directions of the displacements of the ions in the next-nearest neighbor magnetic chains are opposite to each other.
	The transfer operation of $(x+1/2,y+1/2,-z)$ survives, when the lattice is dimerized as shown in Fig. \ref{latdisplacement}(c).
	On the other hand, the lattice is dimerized as shown in Fig. \ref{latdisplacement}(c$'$) when the transfer operation of $(x+1/2,y-1/2,-z)$, which is equivalent to $(x+1/2,y+1/2,-z)$ at room temperature, survives.
	The rectangles drawn by the dashed line in Figs. \ref{latdisplacement}(c), \ref{latdisplacement}(c$'$) and \ref{latdisplacement}(d) clearly show that the $2a \times 2b \times 4c$ superstructure is formed in both the cases.
	Consequently, $C_{s}^{2}$ is one of the possible symmetry in the SP phase of $\alpha '$-NaV$_2$O$_5$.
	This result that two kinds of the displacements of the ions appear below $T_{\rm SP}$ suggests the existence of the domain structure in the SP phase.

	In the SP phase of CuGeO$_3$, the displacement of Cu$^{2+}$ ions along the magnetic chain and the rotation of the O$^{2-}$ ions around the magnetic chain occurred simultaneously.
	The displacements of other ions probably occur in the case of $\alpha '$-NaV$_2$O$_5$.
	When other ions moves so as to break the glide plane in the $ab$ plane, the symmetry of the lattice becomes $C_{1}^{1}$.

	The 62- and 128-cm$^{-1}$ Raman peaks and the continuous Raman band between 130 and 400 cm$^{-1}$ originate probably from the magnetic excitations.
	The 62-cm$^{-1}$ peak slightly shifts to lower frequency and extremely broadens with increasing temperature.
	Fujii {\it et al.} found a SP gap at 9.8 meV (= 78 cm$^{-1}$) in a powder sample by inelastic neutron scattering at a scattering vector $|{\bf Q}| = 1.0$ \AA$^{-1}$.\cite{Fujii96,Yoshihama97}
	Recent inelastic neutron scattering from single crystals revealed a splitting of the magnetic excitations along the $a^*$ axis.
	The lowest and highest energies of the two modes at ${\bf q}=(2.5,0.5,0)$ were observed at about 8.5 meV (= 69 cm$^{-1}$) and 11.3 meV (= 91 cm$^{-1}$).\cite{Yoshihama97b,Yoshihama97c}
	The energy $\Delta$ (= 62 cm$^{-1}$) of this Raman peak is close to that of the low-energy mode at ${\bf q}=(2.5,0.5,0)$, which is transferred to the Brillouin zone center by the formation of the superlattice in the SP phase.
	The energy $\Delta$ of the SP gap was also estimated to be 92 K (= 64 cm$^{-1}$)\cite{Vasilev97} and 85 K  (= 57 cm$^{-1}$)\cite{Schmidt97} by ESR measurements and 98 K (= 68 cm$^{-1}$) by NMR measurement.\cite{Ohama97b}
The first-order Raman scattering from the SP gap excitation is forbidden in pure CuGeO$_3$, since it has inversion symmetry.\cite{Kuroe94b}
It, however, became observable in the Zn- and Si-doped CuGeO$_3$, because the selection rule was broken.\cite{Sekine97,Sekine97b}
On the other hand, the SP gap excitation can be Raman-active in the first-order Raman process even in pure $\alpha '$-NaV$_2$O$_5$, because of the absence of the inversion symmetry in the lattice structure.\cite{Isobe96,Carpy75}
With approaching $T_{\rm SP}$ the weak tail appeared on the low frequency side of the 62-cm$^{-1}$ peak.
There is a possibility that the folding of the Brillouin zone along the $c^*$ axis due to the formation of the $4 \times c$ superlattice induces the observation of the gap excitation at the corresponding wave vectors, whose frequency is almost degenerate against but slightly lower than the 62-cm$^{-1}$ SP gap excitation at ${\bf q}=(2.5,0.5,0)$ because the exchange interaction along the $c$ axis is expected to be very week.

	The asymmetric 128-cm$^{-1}$ peak observed in the $(a, b)$ geometry lies at the bottom of the continuous band appearing between 130 and 400 cm$^{-1}$. 
	The frequency of the Raman peaks in the $(a, b)$ geometry and that of the bottom of the continuous band are almost twice the 62-cm$^{-1}$ peak.
	Thus we ascribe these to the second-order Raman spectrum from the magnetic excitations.
	Appearance of the asymmetric and broad peak at the bottom of the continuous band probably reflects a low-dimensional character of the dispersions of magnetic excitations because the second-order magnetic Raman scattering reflects the density of states of two magnetic excitations.
	We also think that there is another possibility of the formation of the resonant state by attractive interactions between the magnetic excitations.
	The sharp peak due to the two-magnetic-excitation bound state was observed just below 2$\Delta$ in CuGeO$_3$ by Raman scattering,\cite{Kuroe94b} and the peak position crossed $2\Delta$ together with broadening of the halfwidth and weakening of the intensity when Zn or Si atoms were dilutely doped into CuGeO$_3$.\cite{Sekine97,Sekine97b}
	The bound state in pure CuGeO$_3$ changed to a resonant state in the Zn-doped and the Si-doped samples.
	
	The second-order Raman scattering from magnetic excitations was not detected above $T_{\rm SP}$, as shown in Fig. \ref{Rintensity}(a). 
	It, in principle, could be observed even above $T_{\rm SP}$, and the van Hove singurarity coming from the magnetic excitations near the Brillouin zone boundary at 228 cm$^{-1}$ was observed in Cu$_{1-x}$Zn$_x$GeO$_3$, although its intensity decreased rapidly with increasing temperature.\cite{Kuroe96a}
	But the second-order magnetic Raman spectrum below 200 cm$^{-1}$ originates probably from the low-frequency magnetic excitations near the SP gap.
	It is expected that the frequencies of the 128-cm$^{-1}$ peak and the bottom of the broad band decrease together with a decrease of their intensities, as well as the first-order Raman scattering from the SP gap excitation.
	This feature was seen in Fig. \ref{tdependence}(b).
	
	In CuGeO$_3$, the second-order magnetic Raman scattering was observed in the $(c, c)$ geometry,\cite{Kuroe94b,Ogita96} where the incident and the scattered photons were polarized along the 1D antiferromagnetic chain.
	The $(b, b)$ geometry in $\alpha '$-NaV$_{2}$O$_{5}$ corresponds to the  $(c, c)$ geometry in CuGeO$_3$.
	The second-order magnetic Raman spetrum, however, appeared strongly in the $(a, b)$ geometry, although it was weakly observed in the $(b, b)$ geometry, as shown in Fig. \ref{tdependence}(c).
	According to the selection rule the cross-polarization spectrum has a finite intensity in contrast with the case of CuGeO$_3$, because the crystal of $\alpha '$-NaV$_2$O$_5$ is monoclinic or triclinic in the SP phase.
	Since the energy of the electronic excited states of V$^{4+}$ ions is close to that of the incident laser light (514.5 nm),\cite{Golubchik97} we speculate that the cross-polarization spectrum is enhanced by the resonant effect.
	Further studies, however, are needed.

	Golubchik {\it et al.} proposed that the 530-cm$^{-1}$ phonon mode in the $(a,a)$ geometry comes from the Na-O vibration.\cite{Golubchik97}
	But we think that this phonon mode corresponds to the V-O bending $A_{g}$ mode which was reported at 526 cm$^{-1}$ in V$_2$O$_5$.\cite{Abello83}
	Considering the normal coordinate of the 526-cm$^{-1}$ phonon mode in V$_2$O$_5$,\cite{Abello83} one can see that this phonon mode distorts the VO$_5$ pyramid and changes the V-O-V bond angle on the superexchange path along the $b$ axis.
	And therefore it is possible that this phonon mode couples with both the electronic and the magnetic excitations.
	We, however, think that the continuum originates from the $d$-$d$ transition in V$^{4+}$ ions, as proposed by Golubchik {\it et al.},\cite{Golubchik97} because it was strongly observed even at room temperature.
	The amplitude of the electronic Raman scattering may be proportional to the number of V$^{4+}$ ions which is equal to that of Na$^{+}$ ones.
	In V$_2$O$_5$, the continuum was not observed in the $(a,a)$ geometry and the lineshape of the phonon mode at 526 cm$^{-1}$, which corresponds to the 530-cm$^{-1}$ asymmetric Raman peak in $\alpha '$-NaV$_2$O$_5$, is symmetric.\cite{Abello83} 
	These results support our assignment.
	Golubchik {\it et al.} assumed the spherical ($A_{1}$) ground state of the $d$ electron, but this contradicts the $d_{xy}$-like electronic ground state reported by Ohama {\it et al.}\cite{Ohama97}
	The detailed assignment of the $d$-$d$ transition is not clear at the present time.

	The Fano effect originating from the interaction between a phonon and an electronic continuum was observed in Raman spectra of many kinds of solids.\cite{Cerdeira73a,Cerdeira73b,Thomsen88a,Cooper88,Cardona89,Thomsen88b}
	When the discreet excited phonon state decays only to the continuum, the antiresonance point at which the Raman intensity becomes zero appears in the absorption\cite{Fano61} and Raman spectra.\cite{Klein82}
	As shown in Fig. 6, the Raman intensity at the antiresonance point in the sample of $\delta$ = 0 has almost the same as the background.
	But the Raman intensity at the antiresonance point still remains finite, when 1 \% of the Na$^{+}$ ions defect from the sample.
	Moreover, the Raman peak at 531 cm$^{-1}$ in the sample of $\delta$ = 0.1 is almost symmetric, i.e., the Fano effect is very weak in this sample.
	So we need a new theory which gives intermediate spectra between the antiresonance Raman spectra and the usual ones from an noninteracting phonon and a continuous band.
	The theory of Fano effect in the absorption spectrum of a molecule by Nitzan gave the above-mentioned spectra.\cite{Nitzan74}
	He found that the absorption cross section at the antiresonance point remained finite when the phonons decayed to both electronic states and radiations.
	But this theory is not applicable to the Raman process because the phonon cannot directly decay to radiations.
\begin{figure}
\begin{center}
\epsfile{file=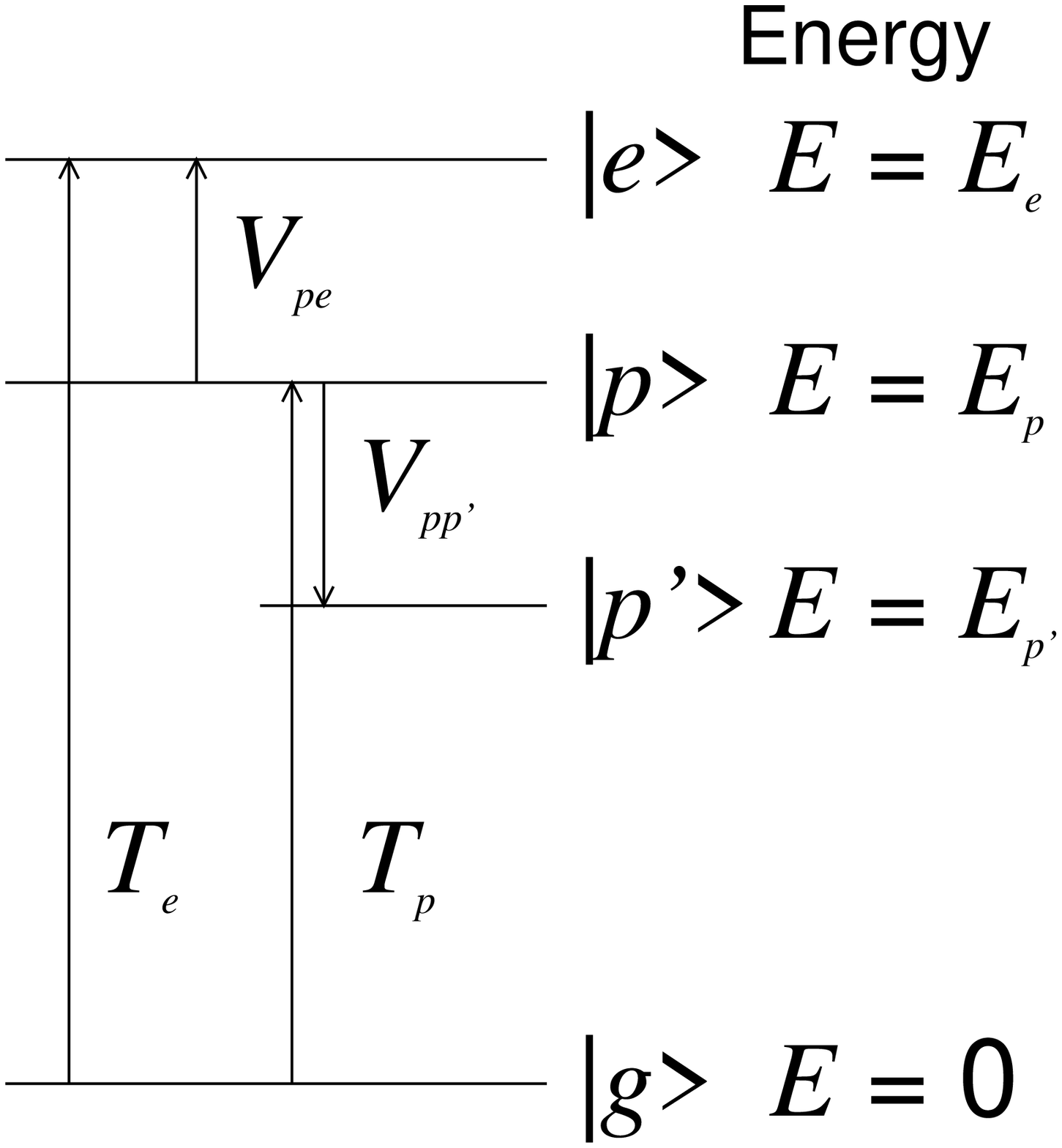,width=0.5\textwidth}
\end{center}
\caption{Energy diagram of Fano effect.}
\label{Fanodiagram}
\end{figure}

	Our theory is based on the Green's function theory of the Fano effect in Raman scattering of solids.\cite{Klein82}
	Figure \ref{Fanodiagram} shows the energy diagram of the system.
	$|g\rangle$ is the unperturbed ground state ($E$ =  $E_{g}$ = 0) of the system.
	The phonon state $|p\rangle$ ($E$ = $E_{p}$) interacts with the continuum $|e\rangle$ ($E$ = $E_{e}$) through the interaction $V_{pe}$.
	We introduce other phonon states $|p'\rangle$, ($E$ = $E_{p'}$ $\neq$ $E_{p}$) so that the phonon $|p\rangle$ can be scattered to another phonon state $|p'\rangle$ by an anharmonic potential and impurities.
	The interaction between $|p\rangle$ and $|p'\rangle$ is defined by $V_{pp'}$.
	Here we assume that $V_{pe}$ and $V_{pp'}$ are real.
	$T_{p}$, $T_{e}$ and $T_{p'}$ denote the Raman tensors of the $|g\rangle \rightarrow |p\rangle$, the $|g\rangle \rightarrow |e\rangle$ and the $|g\rangle \rightarrow |p'\rangle$ Raman processes, respectively.

	The wave function $\psi$ and the Hamiltonian ${\cal H}$ are given by
\begin{equation}
\psi
=
\left(
\begin{array}{c}
|p\rangle\\
|e\rangle\\
|p'\rangle
\end{array}
\right)
\end{equation}
and 
\begin{equation}
{\cal H}
=
\left(
\begin{array}{ccc}
E_{p}   & V_{pe} & V_{pp'} \\
V_{pe}  & E_{e}  & 0       \\
V_{pp'} & 0      & E_{p'}  
\end{array}
\right)
\ , 
\end{equation}
	respectively.
	When $V_{pp'}$ = 0, this Hamiltonian describes the conventional Fano effect.\cite{Klein82}
	Introducing the Green's faction $G(z)$ as
\begin{equation}
G(z) = \lim_{z \rightarrow E + i0^{+}} ({\cal H} - z)^{-1} \ ,
\end{equation}
Raman intensity $l(E)$ is given by 
\begin{equation}
l(E) = {\rm Im} \left\{ \sum_{i,j = e,p,p'} T_{i} \langle i | G | j \rangle T_{j} \right\}\ .
\end{equation}
\begin{full}
	Assuming $T_{p'}$ = 0, we have
\begin{equation}
l(E) = {\rm Im} \left\{ 
\frac{
	T_{e}^{2} \left\{
		(E_{p}-z)(E_{p'}-z)-V_{pp'}^2
	\right\} - 2 T_{p}T_{e}V_{pe}(E_{p'}-z) + T_{p}^2(E_{e}-z)(E_{p'}-z)
}{
	(E_{p}-z)(E_{e}-z)(E_{p'}-z) - V_{pe}^{2} (E_{p'}-z) - V_{pp'}^{2}(E_{e}-z)
}
\right\}\ .
\end{equation}
	To treat the states $|e\rangle$ and $|p'\rangle$ as continua, the terms of $(E_{j}-z)$, where $j$ = $e$ and $p'$, may be replaced by $(-R_{j}(E)+i\pi \rho_{j}(E))^{-1}$.\cite{Klein82}
	Here $\rho_{e}(E)$ and $\rho_{p'}(E)$ are respectively the densities of states of the states $|e\rangle$ and $|p'\rangle$.
	When $\rho_{e}(E)$ and $\rho_{p'}(E)$ vary slowly against $E$, we can neglect the terms of $R_{e}(E)$ and $R_{p'}(E)$, which are respectively the Hilbert transformations of $\rho_{e}(E)$ and $\rho_{p'}(E)$.
	We obtain
\begin{equation}
l(E) = 
\frac{\Gamma_{e}}{\Gamma}
T_{e}^{2} \pi \rho_{e}(E)
\frac{(E - E_{p} +\frac{T_{p}}{T_{e}} V_{pe})^{2}}
{(E - E_{p})^{2} + \Gamma^{2}}
+
\frac{\Gamma_{p'}}{\Gamma} T_{e}^2
\left[
\pi \rho_{e}(E)
\frac{(E  - E_{p} + \frac{T_{p}}{T_{e}} V_{pe})^{2}+\Gamma^{2}}
{(E  - E_{p})^{2} + \Gamma^2}
\right.
+ \left( \frac{T_{p}}{T_{e}} \right)^2
\left.
\frac{\Gamma}
{(E - E_{p})^{2} + \Gamma^2}
\right]
\ , 
\label{KuroeFano}
\end{equation}
where $\Gamma_{e} = \pi \rho_{e}(E) V_{pe}^2$ and $\Gamma_{p'} = \pi \rho_{p'}(E) V_{pp'}^{2}$, denoting the halfwidths which are equal to the inverse of the decay rate of $|p\rangle \rightarrow |e\rangle$ and $|p\rangle \rightarrow |p'\rangle$ processes, respectively.
\end{full}
And $\Gamma$ (= $\Gamma_{e} + \Gamma_{p'}$) gives the total halfwidth which corresponds to FWHM of the phonon Raman peak.
We notice here that $\Gamma_{e}/\Gamma$ and $\Gamma_{p'}/\Gamma$ give the relative decay rates which are equal to the transition probabilities of the decay channel $|p\rangle \rightarrow |e\rangle$ and $|p\rangle \rightarrow |p'\rangle$, respectively.
	One can easily see that eq. (\ref{KuroeFano}) leads to the conventional Fano equation when $\Gamma_{p'}/\Gamma_{e} \ll 1$.
	On the other hand, when $\Gamma_{p'}/\Gamma_{e} \gg 1$, it gives a phonon Raman scattering with a Lorentzian lineshape at $E_{p}$, with a coupling coefficient $T_{p}$ and a halfwidth $\Gamma_{p'}$, which is superimposed on the continuum $\pi\rho_{e}(E)$.

Since the halfwidths around $E_{p}$ are important for the Fano effect, the energy dependence of $T_{p}$, $T_{e}$, $\Gamma_{e}$ and $\Gamma_{p'}$ are negligible.
	Finally the simple form of Raman spectrum $I(\tilde E)$ is given as
\begin{eqnarray}
\nonumber
I(\tilde E) & = &
\frac{l(\tilde E)}{\pi \rho_{e}(\tilde E) T_{e}^{2}}
\\ & = & 
\nonumber
\frac{(\tilde E + T_{ep}V_{pe})^2 
 + \Gamma_{p'}(\Gamma_{e} + \Gamma_{p'} + (T_{ep}V_{pe})^{2}/\Gamma_{e})}
{\tilde E^{2} + (\Gamma_{e} + \Gamma_{p'})^2}
 \ ,\\ & &
\label{Kuroeeq}
\end{eqnarray}
where $T_{pe} = T_{p}/T_{e}$ and $\tilde E = E - E_{p}$.
When $T_{e}$ = 1, $\pi \rho_{e}(\tilde E) = 1$ and $V_{pe} > 0$, we obtain
\begin{equation}
I(\tilde E_{\rm min})  =  \Gamma_{p'}/\Gamma \  {\rm at }\  \tilde E_{\rm min}  =  -T_{ep}\Gamma / V_{pe} \ , 
\end{equation}
and
\begin{equation}
I(\tilde E_{\rm max})  =  1 + T_{ep}^2 / \Gamma \ {\rm at}\ \tilde E_{\rm max}  =  V_{pe}\Gamma / T_{ep} .
\end{equation}

\begin{figure}[b]
\begin{center}
\epsfile{file=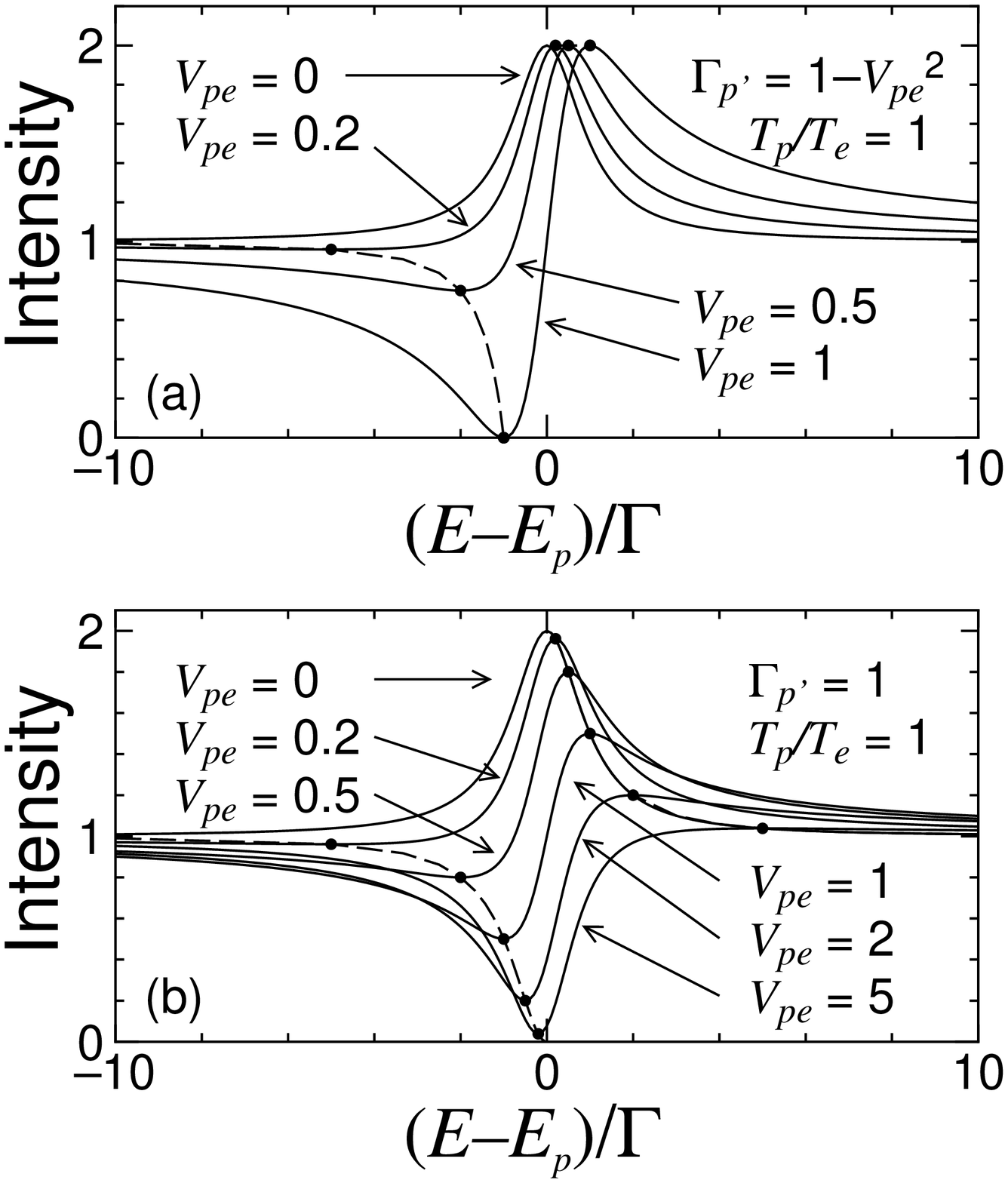,width=0.5\textwidth}
\end{center}
\caption{Calculated Raman spectra with various $V_{pe}$.
The inverse decay rate of $|p\rangle \rightarrow |p'\rangle$ process and the total halfwidth of the phonon mode are fixed to be unity in (a) and (b), respectively.
The ordinates are divided by the total halfwidth of the phonon mode. 
The Raman tensors of the phonon Raman process and that of the electronic Raman process are fixed to be unity. The dashed curves with solid circles are the traces of the maximum and the minimum intensities of Raman spectra.}
\label{energy2}
\end{figure}

Figure \ref{energy2}(a) shows the calculated Raman spectra under the condition of a constant $\Gamma$.
When $V_{pe} = 0$, the Raman spectrum is a sum of a Lorentzian and a continuum.
The asymmetric peak shifts to the higher-energy side with increasing $V_{pe}$, and its Raman intensity at the peak position does not depend on $V_{pe}$.
On the other hand, the antiresonance energy approaches $-T_{ep}\Gamma/V_{pe}$ and the Raman intensity at the antiresonance point approaches zero.
When the transition probability of the $|p\rangle \rightarrow |e\rangle$ channel decreases, the effect of the antiresonance weakens.

Figure \ref{energy2}(b) shows the calculated Raman spectra under the condition of a constant $\Gamma_{p'}$.
The asymmetric peak shifts to the higher-energy side and the peak intensity decreases with increasing $V_{pe}$.
On the other hand, both the antiresonance energy and the Raman intensity at the antiresonance point approach zero with increasing $V_{pe}$.

	Let us apply our theory to the observed data.
	We assume that the Fano effect is very weak in the sample of $\delta = 0.1$, because the lineshape of the Raman peak at 530 cm$^{-1}$ in this sample is almost symmetric.
	The density of states of the continuous band $(1-\delta)\rho(\omega)$ in the sample of $\delta = 0.1$ can be represented phenomenologically by the sum of two Lorentz functions at 485 and 566 cm$^{-1}$ which is denoted by the solid line in Fig. \ref{fanoeffect}.
	This curve represents the Raman spectrum between 450 and 800 cm$^{-1}$ well except for the 530-cm$^{-1}$ peak.
	Other small discrepancies between this curve and the observed Raman spectra will be stated later.

\begin{fulltable}
\begin{fulltabular}{*{8}{c}}
$\delta$ & $\omega_0$ [cm$^{-1}$] & $T_{ep}V_{pe}$ [cm$^{-1}$]& $\Gamma_{p'}$ [cm$^{-1}$] & $\Gamma_{e}$ [cm$^{-1}$]& $\Gamma$ [cm$^{-1}$]& $\sqrt{\frac{(T_{ep}V_{pe})^{2}(1-\delta)}{\Gamma_{e}}}$ & $\sqrt{\frac{\Gamma_{e}}{1-\delta}}$ \\
\hline
0    & 529   & 6.4  & 2.5  & 2.2    & 4.7  & 4.3  & 1.5  \\
0.01 & 529   & 5.1  & 3.2  & 1.5    & 4.7  & 4.1  & 1.2  \\
0.1  & 530   & 1.1  & 6.3  & 0.058  & 6.3  & 3.9  & 0.25 \\
\hline
\end{fulltabular}
\caption{Parameters obtained from the fitting.}
\label{fanoprm}
\end{fulltable}
	We fitted the Raman spectra in the samples of $\delta$ = 0, 0.01 and 0.1 to the function of $(1-\delta)\rho(\omega)I(\omega) + {\rm background}$.
	Here we used $E_{p}$ (= $\hbar \omega_0$), $T_{ep}V_{pe}$, $\Gamma_{e}$ and $\Gamma_{p'}$ as parameters and assumed the background as a function of $a \omega + b$.
	As shown in Fig. \ref{fanoeffect}, the calculated curves (dotted curves) reproduced the observed data well.
	We obtain the parameters $\omega_{0}$, $T_{ep}V_{pe}$, $\Gamma_{e}$, $\Gamma_{p'}$, $\Gamma$, $\sqrt{(T_{ep}V_{pe})^2(1-\delta)/\Gamma_{e}}$ ($\propto T_{ep}$) and $\sqrt{\Gamma_{e}/(1-\delta)}$ ($\propto V_{pe}$) as listed in Table \ref{fanoprm}.
	$V_{pe}$ decreases with increasing defects of Na$^{+}$ ions, indicating that the interaction between the phonon state and the continuum rapidly decreases.
	$\Gamma_{e}/\Gamma$ in the sample of $\delta = 0.1$ is much smaller than those in the samples of $\delta$ = 0 and 0.01 and it is confirmed that the Fano effect in the sample of $\delta$ = 0.1 is very weak.
	On the other hand, $\Gamma_{p'}$ and $\Gamma$ increase with increasing defects of Na$^{+}$ ions.
	This result and the fact that the FWHMs of the 690- and the 965-cm$^{-1}$ phonon modes also increase with increasing $\delta$, as shown in Fig. \ref{defectedsamples} and Table \ref{fanoprm}, because these phonons are scattered to the local phonons at the defect sites.
	The bare phonon frequency $\omega_0$ (= 529 cm$^{-1}$) in the sample of $\delta$ = 0 and 0.01 are close to that in the sample of $\omega_0$ (= 530 cm$^{-1}$) in the sample of $\delta$ = 0.1.
	$T_{ep}$ decreases with increasing defects of Na$^{+}$ ions, which may be caused by the decrease of Raman tensor of the phonon.

	As shown in Fig. \ref{fanoeffect}, the small asymmetric peak at 455 cm$^{-1}$ and the dip at 431 cm$^{-1}$ were observed.
	The dip at 431 cm$^{-1}$ was also observed even at room temperature.
	The intensity of the asymmetric peak at 455 cm$^{-1}$ weakens and the depth of the dip increases with increasing the defect of Na$^{+}$ ions.
	The dip at 431 cm$^{-1}$ and the asymmetric peak at 455 cm$^{-1}$ may also be interpreted by the Fano resonance.

\section{Conclusion}
	We measured polarized Raman spectra in $\alpha '$-Na$_{1-\delta}$V$_2$O$_5$ ($\delta$ = 0, 0.01 and 0.1).
	When $\delta$ = 0, we observed six, seven and three Raman peaks at room temperature in the $(a,a)$, $(b,b)$ and $(a,b)$ geometries, respectively.
	By means of the group-theory analysis, we assigned six peaks as $A_1$ phonon modes and three peaks as $A_2$ ones.
	The 418-cm$^{-1}$ peak at room temperature in the $(b,b)$ geometry has a broader halfwidth than any other peaks.
	At 15 K this peak shifted to 425 cm$^{-1}$ and a small shoulder at 393 cm$^{-1}$ appeared.
	It is not a usual phonon mode.
	Moreover, we observed five new peaks at 62, 102, 128, 646 and 944 cm$^{-1}$ and a broad Raman band between 130 and 400 cm$^{-1}$ below $T_{\rm SP}$.
	We assigned the peaks at 102 and 944 cm$^{-1}$ to the folded phonon modes induced by the formation of the superlattice structure in the SP phase.
	The peak at 62 cm$^{-1}$ comes from the first-order Raman scattering from the spin gap excitation, and the 128-cm$^{-1}$ peak and the new broad band are assigned the second-order magnetic Raman scattering.
	The temperature dependence of the intensities and that of the halfwidths of these new peaks were measured.
	We obtained that $T_{\rm SP}$ = 31 K from the intensities of the new peaks in the $(a,a)$ and $(a,b)$ geometries.
	On the other hand, the new peaks in the $(b,b)$ geometry disappeared above 25 K.
	This originated from the local temperature rise due to the incident laser.
	Since the folded phonon mode at 944 cm$^{-1}$ was observed in the $(a,a)$ and $(a,b)$ geometries, the lattice in the SP phase belongs to monoclinic structure of $C_{s}^{2}(Pn)$ or triclinic structure of $C_{1}^{1}(P1)$.

	We studied the effects of the Na$^{+}$-ion deficiency.
	With increasing defects of Na$^{+}$ ions, the tail of the incident laser line which came from the quasielastic scattering or the single particle excitation of carriers increased.
	The folded phonon modes at 646 and 944 cm$^{-1}$ broadened and weakened in the sample of $\delta$ = 0.01, and they were not observed in the sample of $\delta$ = 0.1.
	The defects of Na$^{+}$ ions suppressed the SP transition.

	At both the room temperature and 15 K, we observed a broad band around 500 cm$^{-1}$ superimposed on an asymmetric peak at 531 cm$^{-1}$ in the $(a,a)$ geometry.
	We attributed this lineshape to the Fano effect which is owing to the interaction between the V-O bending mode and the $d$-$d$ electric transition in V$^{4+}$ ions.
	When the Na$^{+}$ ions defected from the lattice, this peak became symmetric and the antiresonance of Raman spectrum weakened.
	We introduced the effect of the finite lifetime of the phonon into the Green's function theory of the Fano effect in order to discuss this phenomenon.
	Our theory reproduced well the Na$^{+}$-ion deficiency dependence of Raman spectra in this system.

\section*{Acknowledgements}
	This study was supported in part by a Grant-in-Aid for Science Research from the Ministry of Education, Science, Sports and Culture of Japan and by the Research Fellowships of Japan Society for the Promotion of Science for Young Scientists (HK).

\end{document}